\pdfoutput=1

\input{aipcheck}

\documentclass[
    ,final            
    ,numberedheadings 
  ]
  {aipproc}

\layoutstyle{6x9}

\begin{document}

\title
{Nonlinear Dynamics of Bose-Einstein Condensates with Long-Range Interactions}


\author{G. Wunner}{
  address={Institut f\"ur Theoretische Physik 1, Universit\"at Stuttgart, 70550 Stuttgart, Germany}
}
\author{H. Cartarius}{
  address={Institut f\"ur Theoretische Physik 1, Universit\"at Stuttgart, 70550 Stuttgart, Germany}
}
\author{T.  Fab\v ci\v c}{
  address={Institut f\"ur Theoretische Physik 1, Universit\"at Stuttgart, 70550 Stuttgart, Germany}
}
\author{P. K\"oberle}{
  address={Institut f\"ur Theoretische Physik 1, Universit\"at Stuttgart, 70550 Stuttgart, Germany}
}
\author{J. Main}{
  address={Institut f\"ur Theoretische Physik 1, Universit\"at Stuttgart, 70550 Stuttgart, Germany}
}
\author{T. Schwidder}{
  address={Institut f\"ur Theoretische Physik 1, Universit\"at Stuttgart, 70550 Stuttgart, Germany}
}

%
%

\begin{abstract}
The motto of this paper is: Let's face Bose-Einstein condensation
through nonlinear dynamics. We do this by choosing variational forms
of the condensate wave functions (of given symmetry classes),
which convert the Bose-Einstein condensates via the time-dependent 
Gross-Pitaevskii equation into Hamiltonian systems that can be studied using the
methods of nonlinear dynamics. We consider in particular cold quantum
gases where  long-range interactions between the neutral atoms
are present, in addition to the conventional short-range contact 
interaction, viz. gravity-like interactions, and dipole-dipole interactions.
The results obtained serve as a useful guide in the search for 
nonlinear dynamics effects in numerically exact quantum calculations 
for Bose-Einstein
condensates. A main result is the prediction of the existence of
stable islands as well as chaotic regions for excited states of
dipolar condensates, which could be checked experimentally.
\end{abstract}

\maketitle

\section{Introduction}

 At sufficiently low temperatures
a condensate of weakly interacting bosons can be represented by a single wave function 
whose dynamics obeys the Gross-Pitaevskii equation \cite{gross61,pitaevskii61}. The Gross-Pitaevskii equation can be thought of as the Hartree equation for the ground state
of $N$ interacting identical bosons, all occupying the same single-particle orbital $\psi$.
 Because of its nonlinearity
the equation exhibits features not familiar from ordinary Schr\"odinger equations of quantum mechanics.
For example, Huepe et al. \cite{hue99,hue03} demonstrated that for Bose-Einstein condensates
with attractive contact interaction, described by a negative  $s$-wave scattering length, {\em bifurcations} 
of the stationary solutions of the Gross-Pitaevskii equation appear, and  determined both the stable
(elliptic) and the unstable (hyperbolic) branches of the solutions. The bifurcation points correspond to critical
particle numbers, above which, for given strength of the attractive interaction, collapse of the condensate sets in. In Bose-Einstein condensates of $^7$Li \cite{sacket99,gerton00}
and $^{85}$Rb atoms \cite{donley01,roberts01} these collapses were  experimentally observed.
 
In those condensates the short-range contact interaction is the only interaction to be considered. In Bose-Einstein
condensates of {\em dipolar} gases \cite{santos00,baranov02,goral02a,goral02b,giovanazzi03} also a long-range dipole-dipole interaction is present.  
Alternatively, following a proposal by O'Dell et al. \cite{dell00,giova01},
by using a combination
of 6 triads of appropriately tuned laser light condensates can be produced
in which an attractive long-range gravity-like $1/r$ interaction is present.
These types of condensates offer the 
opportunity to study degenerate quantum gases with adjustable long-range {\em and} short-range interactions. While the experimental realization of condensates
with gravity-like interaction lies still in the future, the
achievement of Bose-Einstein condensation in a gas of chromium atoms  \cite{griesmaier05}, with a large dipole moment, has opened the
way to promising experiments on dipolar gases  \cite{stuhler05}, which could show a wealth of novel phenomena \cite{giovanazzi02,santos03,li04,dell04}. In particular,
the experimental observation of the collapse of dipolar quantum gases has been reported  \cite{koch08} which occurs when 
the contact interaction is reduced, for a given particle number, below some critical value using a Feshbach resonance.

In this experimental situation it is most timely and appropriate to extend the investigations of the nonlinearity effects of the
Gross-Pitaevskii equation to quantum gases in which both the contact interaction 
and a long-range interaction is active, and this is the topic of the present paper.

\section{Scaling Properties of the Gross-Pitaevskii Equations with   Long-Range Interactions}

\subsection{Gravity-like interaction, isotropic trap}\label{scaling1/r}

For an additional gravity-like long-range interaction $ 
V_u(\vec r, \vec r^\prime \,) = - {u}/{|\vec r - \vec r^\prime |}$
the time-dependent Gross-Pitaevskii equation for the orbital $\psi$ reads
\begin{equation}\label{HFat}
\Big[ \hskip -1mm - \Delta + \gamma^2 r^2 + N 8 \pi \frac{a}{a_u} |\psi(\vec r, t)|^2 
- 2 N  \hskip -1mm  \int \frac{|\psi({\vec r\,}^\prime,t)|^2}{|{\vec r}- {\vec r\,}^\prime |}
d^3 {\vec r\,}^\prime \Big] \; \psi(\vec r, t) = i  \frac{\partial}{\partial t}  \, \psi(\vec r, t) \, , 
\end{equation}
where we have used \cite{Pap07} the ``Bohr radius'' $a_u = {\hbar^2}/{(m u)}$, the ``Rydberg energy''  
$E_u =  {\hbar^2}/{(2m a_u^2)}$, and the Rydberg time $\hbar/E_u$ as natural units
of length, energy, and time, respectively, to bring the equation in dimensionless form. 
Furthermore, in  (\ref{HFat}) $a$ is the  $s$-wave scattering length, which  characterizes the strength
of the contact interaction
$V_s = \delta({\vec r}- {\vec r\,}^\prime)\, 4 \pi a \hbar^2/m $,
$N$ is the particle number, and $\gamma = \hbar \omega_0/(2E_u)$ is the dimensionless trap frequency.
It can be shown \cite{Pap07} that the solutions of  (\ref{HFat}) do not depend on all these
three physical quantities but only on the two relevant parameters $\gamma/N^2$ and $N^2 a/a_u$. Thus
one has, e.g., for the mean-field energy  $E (N, N^2 a^\ast/a_u, \gamma/N^2)/N^3 = \,E(N=1, a/a_u, \gamma)$, with $ N^2 a^\ast/a_u =   a/a_u$.   

\subsection{Dipolar interaction, axisymmetric trap}\label{scaling}

In Bose condensates of $^{52}$Cr atoms \cite{griesmaier05}, which possess a large magnetic moment of $\mu = 6 \mu_{\rm B}$, the long-range dipole-dipole interaction 
\begin{equation}
V_{\rm dd}({\vec r}, {\vec r\,'}) =\frac{\mu_0 \mu^2}{4 \pi} \,\frac{1-3\cos^2 \theta'}{|{\vec r}-{\vec r}\,'|^3} \nonumber 
\end{equation}
must also be considered. Defining the dipole length by $a_{\rm d} = {\mu_0 \mu^2 m}/{(2\pi \hbar^2)}$,
and using as unit of energy $E_{\rm d} = \hbar^2/(2m a_{\rm d}^2)$, of frequency $\omega_{\rm d} = 
2 E_{\rm d}/\hbar$ and of time $\hbar/E_{\rm d}$, one obtains the Gross-Pitaevskii equation for dipolar 
gases in axisymmetric traps in dimensionless form
\begin{eqnarray}\label{gpedd}{\Big[} \hskip -1mm - \Delta + \gamma_{\rho}^2 \rho^2 + \gamma_z^2 z^2
  +  N 8 \pi \frac{a}{a_{\rm d}} |\psi({\vec r},t)|^2
\hspace*{-2mm} &+& \hspace*{-2mm}
  N  \hskip -1mm  \int |\psi({{\vec r}\,}^\prime,t)|^2 
\frac{(1 - 3 \cos^2
  \vartheta^\prime)}{|{{\vec r}}- {{\vec r}\,}^\prime |^3}
 d^3 {{\vec r}\,}^\prime
 \Big] \; \psi({\vec r},t) \nonumber \\  &=& \hspace*{-2mm}i \frac{\partial}{\partial t} \, \psi({\vec r},t) \, .
\end{eqnarray}
The physical parameters characterizing the condensates are the particle number $N$, the scattering length
$a/a_{\rm d}$ and the trap frequencies $\gamma_\rho$ and $\gamma_z$ perpendicular to and along the
direction of alignment of the dipoles (alternatively, the geometric mean 
(${\bar{\gamma}} = \gamma_\rho^{2/3} \gamma_z^{1/3}$ and the aspect ratio $\lambda = \gamma_z/\gamma_\rho$ 
is used). However, a closer inspection of the scaling properties of (\ref{gpedd}) reveals \cite{Koeb08}
that the solutions depend only on three parameters, viz. 
 $N^2 {\bar{\gamma}}, ~\lambda, ~a/a_{\rm d}$. For the mean-field energy, e.g., the scaling law
reads 
$E (N, a/a_{\rm d}, N^2{\bar{\gamma}}^\ast, \lambda)  \,=\,E(N=1, a/a_{\rm d}, 
{\bar{\gamma}}, \lambda)\, /N$, with $ N^2{\bar{\gamma}}^\ast = {\bar{\gamma}}$. 

\section{Quantum Results: Solutions of the Stationary Gross-Pitaevskii Equations}

For the $1/r$ interaction (monopolar quantum gases) we have solved  \cite{Pap07} the stationary Gross-Pitaevskii
equation both variationally, using an isotropic Gaussian-type orbital  $\psi = A \exp(-k^2 r^2/2)$,
and numerically accurate, by outward integration of the nonlinear Schr\"odinger equation.
For the dipole-dipole interaction (dipolar quantum gases) we have performed a variational calculation  \cite{Koeb08}
using an axisymmetric Gaussian-type orbital  $\psi = A \exp(-k_{\rho}^2 {\rho}^2/2-k_z^2 z^2/2)$.

\begin{figure}\label{bifurcationmonopolar1}
\caption{Bifurcations of the particle-number scaled chemical potential. Left: $1/r$ interaction, for different trap 
frequencies $\gamma$ (in units of $N^2$); solid curves: accurate numerical calculation, dashed curves: variational calculation. Right:
dipole-dipole interaction, variational results for the geometric mean trap frequency 
$N^2 \bar \gamma = 3.4 \times 10^4$ used in the experiments of Koch et al. \cite{koch08}
and different values of the trap aspect ratio $\lambda$.}
\includegraphics[width=\textwidth]{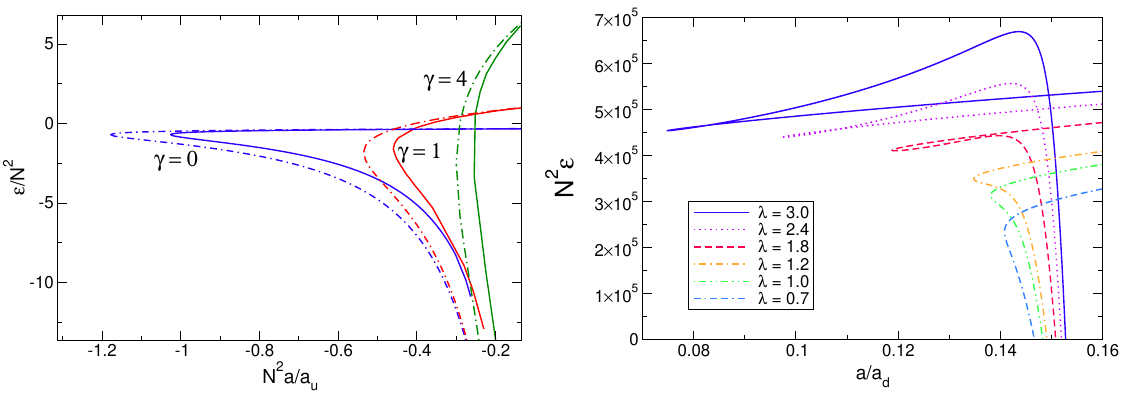}
\end{figure}

Fig.~\ref{bifurcationmonopolar1} shows the results for  the chemical potential (eigenvalue of the stationary Gross-Pitaevskii 
equation) for the two interactions, plotted as a function of the scattering length. As can be seen, below a critical scattering
length no stationary solutions exist, while two stationary solutions are born at the critical scattering length in a tangent 
bifurcation. At the bifurcation point the chemical potential, the mean-field energy, and the wave functions of the 
two branches of solution are identical. Such behavior is obviously a consequence of the nonlinearity of the underlying
Schr\"odinger equation, and is reminiscent of exceptional points \cite{Hei99,Kato66} discussed so far only in the context of  
open quantum systems with non-Hermitian Hamiltonians (see Ref. \cite{Car08a} for references). In fact, a closer inspection shows \cite{Car08a} that the bifurcation
points can be identified as exceptional points: traversing circles around them in the 
complex-extended parameter plane, the
eigenvalues are permuted, which is a clear signature of exceptional points. 

\section{Nonlinear Dynamics of Bose-Einstein Condensates with Atomic Long-Range Interactions}
Starting point of accurate numerical calculations are the time-dependent Gross-Pitaevskii equations 
(\ref{HFat}) and  (\ref{gpedd}). For variational calculations one makes use of the
fact that these equations follow from the variational principle
$||i \phi(t) -H \psi(t)||^2 ={\rm min}$, where the variation is performed with respect to $\phi$, and finally $\phi$ is set equal to $\dot \psi$. Using
a complex parametrization of the trial wave function  $\psi(t) = 
\chi({\vec \lambda}(t))$, the variation leads to equations of motion for the parameters $ {\vec \lambda}$ (cf. \cite{Fab07})
\begin{equation} \label{vareq}\textstyle{
 \left\langle \frac{\partial \psi}{\partial {\vec \lambda}}\Big|i \dot \psi - H \psi \right\rangle
 = 0  \leftrightarrow  
K \dot {\vec \lambda} = -i {\vec h} \hbox{~with~} K = \left\langle \frac{\partial \psi}{ \partial{{\vec \lambda}}}
 \Big|\frac{\partial \psi}{\partial{{\vec \lambda}}} \right\rangle, 
  {\vec h} = \left\langle \frac{\partial \psi}
 { \partial{{\vec \lambda}}}
 \Big|H \Big|\psi \right\rangle}\,.
\end{equation}

\subsection{Time evolution of condensates with $1/r$ interaction, variational and exact}\label{1/r}
\begin{figure}\label{phaseportraits}
\caption{Phase portraits of the dynamics of the complex width function $A(t)$ associated with the 
Hamiltonian (\ref{varham}) for attractive $1/r$ interaction for different values of the
scattering length $a$, measured in units of $a_u$.
Left: $a = -1 > a_{\rm cr}$: two stationary states appear as fixed points; 
middle:  $a = -1.18 =  a_{\rm cr}$: coalescence of the fixed points; right:
$ a = -1.3 < a_{\rm cr}$: no stationary solutions exist.}
\includegraphics[width=\textwidth]{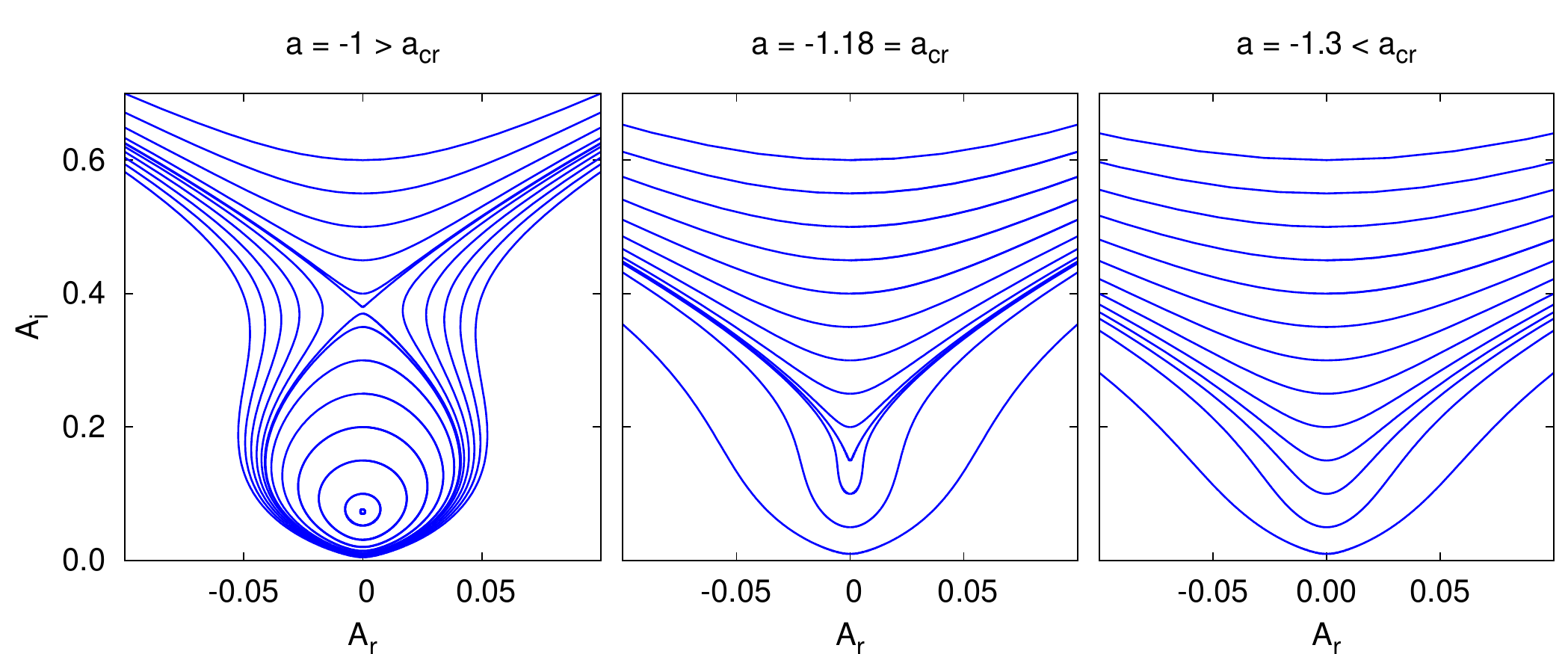}
\end{figure}

For simplicity we consider the case of selftrapping, with no external trap.
As one can convince oneself, the results can be easily generalized
to the case where an external radially symmetric trap potential is
present. 
We choose a Gaussian trial wave function $\psi(r,t) = \exp\{ {i} [A(t) r^2 + \gamma(t)]\}$, where $A$ and $\gamma$ are complex functions of time, whose  
equations of motion follow from (\ref{vareq}). Decomposing $A$ into real and imaginary parts,  $A = A_r + i A_i $, and replacing them by  two other
dynamical quantities \cite{Bro89,Car08b}, $q = \sqrt{3/(4 A_i)} \equiv \sqrt{\langle r^2 \rangle}$, $ p = A_r\sqrt{3/A_i}$, converts those equations into
the canonical  equations of motion for $p$ and $q$ that follow from the 
 Hamiltonian
\begin{equation}\label{varham}
 E = H(q,p) = T+V = p^2+\frac{9}{4q^2}
 +\frac{3 \sqrt{3}a}{2\sqrt{\pi}q^3} -\frac{\sqrt{3}}{\sqrt{\pi}q}. \nonumber
\end{equation}
In this way the Gross-Pitaevskii equation is mapped onto the Hamiltonian of a one-dimensional
classical autonomous system with a nonlinear potential $V(q)$. 
The potential has no extremum for $a < a_{\rm cr} = -3\pi/8 \approx  -1.18$, possesses
a saddle point for $a = a_{\rm cr}$, and 
a maximum and a minimum for $a > a_{\rm cr}$. The critical scattering length
corresponds to the bifurcation point in the variational calculation.
For different values of $a$ (in units of $a_u$) phase portraits of trajectories moving according to the Hamiltonian (\ref{varham}) are shown in Fig.~\ref{phaseportraits}.

The linear stability analysis of both the variational and the exact quantum 
stationary solutions proves \cite{Car08b} that the state corresponding to the elliptical
fixed-point indeed is dynamically stable (small perturbations of the
state show oscillating behavior), while the stationary state 
corresponding to the hyperbolic fixed-point is dynamically unstable 
(exponential growth of small perturbations).

This behavior is recovered in exact numerical solutions of the 
time-dependent Gross-Pitaevskii equation with $1/r$ interaction, but also 
new features emerge. The solution is carried out \cite{Car08b} using
the split-operator technique and fast Fourier transforms. To investigate
the behavior of condensate wave functions in the vicinity of the
exact numerical stable and unstable stationary states, we consider 
condensates which are obtained by deforming the stationary states
by $\psi(r)  = f\cdot \psi_{\pm}(r\cdot f^{2/3})$, where $f$ is a 
numerical stretching factor (this choice of the perturbation does not 
affect the norm of the state).

\begin{figure}\label{wavefunctions}
\caption{Time evolution, for attractive $1/r$ interaction, of the root-mean-square widths of the condensate wave functions
 in the vicinity
of the unstable (panels a), b), c)) and the stable stationary (panels d), e)) 
state. a): Scaled scattering length (in units of $a_{\rm d}$) a = -1.0, stretching factor
f= 1.00; b): a = -0.85, f = -1.001; c): a = -0.85, f = 0.99; d): a = -0.85,
f =1.25,  and e): a = -0.85, f = 1.01. 
}
\includegraphics[width=1 \textwidth]{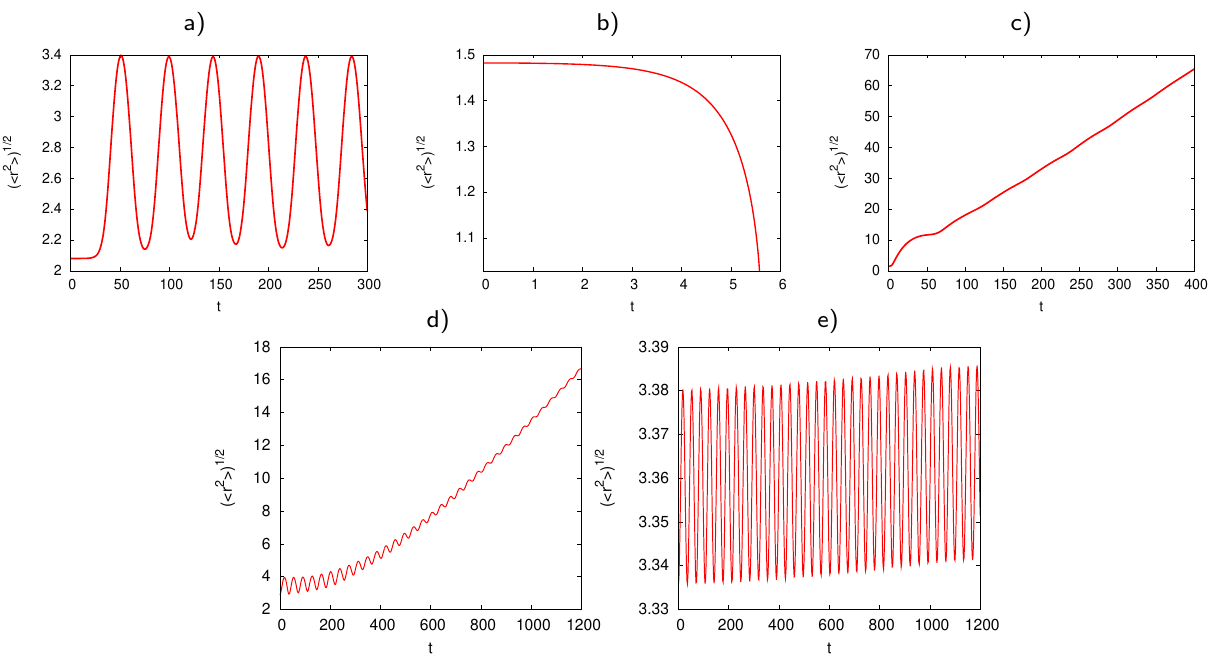}
\end{figure}

In Fig.~\ref{wavefunctions} we show examples of the exact BEC dynamics
in the vicinity of the unstable and stable stationary states. 
In Fig.~\ref{wavefunctions} a) we start the time evolution with
the numerical solution for the unstable stationary state (in the classical picture this
corresponds to the trajectory starting at the hyperbolic fixed point,
see left part of Fig.~\ref{phaseportraits}). Because of unavoidable 
numerical deviations from the theoretically exact unstable state, the 
 wave function determined numerically
 is stationary only for some time but then
begins to oscillate. Obviously we have started with a state which
in the variational picture would be located in the elliptical domain
close to the hyperbolic fixed point. Note, however, that the
oscillation is not strictly periodic. By contrast, in 
Fig.~\ref{wavefunctions} b), where the time evolution starts 
with the unstable stationary state stretched by a factor of $f = 1.001$,
as time proceeds the wave function contracts towards the origin, 
and the condensate collapses. In the variational picture this corresponds
to a trajectory initially close to the hyperbolic fixed point but
located on the hyperbolic side. Note that in a realistic experimental
situtation during the  collapse further mechanisms have to be
taken into account, such as inelastic collisions. The inclusion of such
mechanisms, however, clearly goes beyond the scope of the present paper.

Figure~\ref{wavefunctions} c) displays a behavior not present in the variational
analysis. We start again in the vicinity of the unstable
stationary state ($f = 0.99$) and find that the width of the
condensate gradually grows and grows. An inspection of the 
wave function on a logarithmic scale shows \cite{Car08b} that
wave function amplitude builds up at large distances from 
the origin, giving rise to this behavior. Finally,
Fig.~\ref{wavefunctions} d), e)  show examples for the quantum mechanical
time evolution of condensates in the vicinity of the stable 
ground state. For a large stretching factor (panel d)) 
the condensate is found to oscillate and to expand, while
for a small stretching factor (panel e)) we find the quasiperiodic oscillations
that we would expect from the variational analysis. This demonstrates
that the variational nonlinear dynamics approach is capable of 
predicting essential features of the exact quantum mechanical 
time behavior of the condensates, but that the quantum mechanical
behavior is even richer.

\subsection{Dynamics of condensates with dipolar interaction, variational}
\begin{figure}\label{vnumplot2}
\caption{Potential $V(q_\rho, q_z)$ in the Hamiltonian (\ref{hamiltoniandd}) for
dipole-dipole long-range interaction.}
\includegraphics[width=0.55\textwidth]{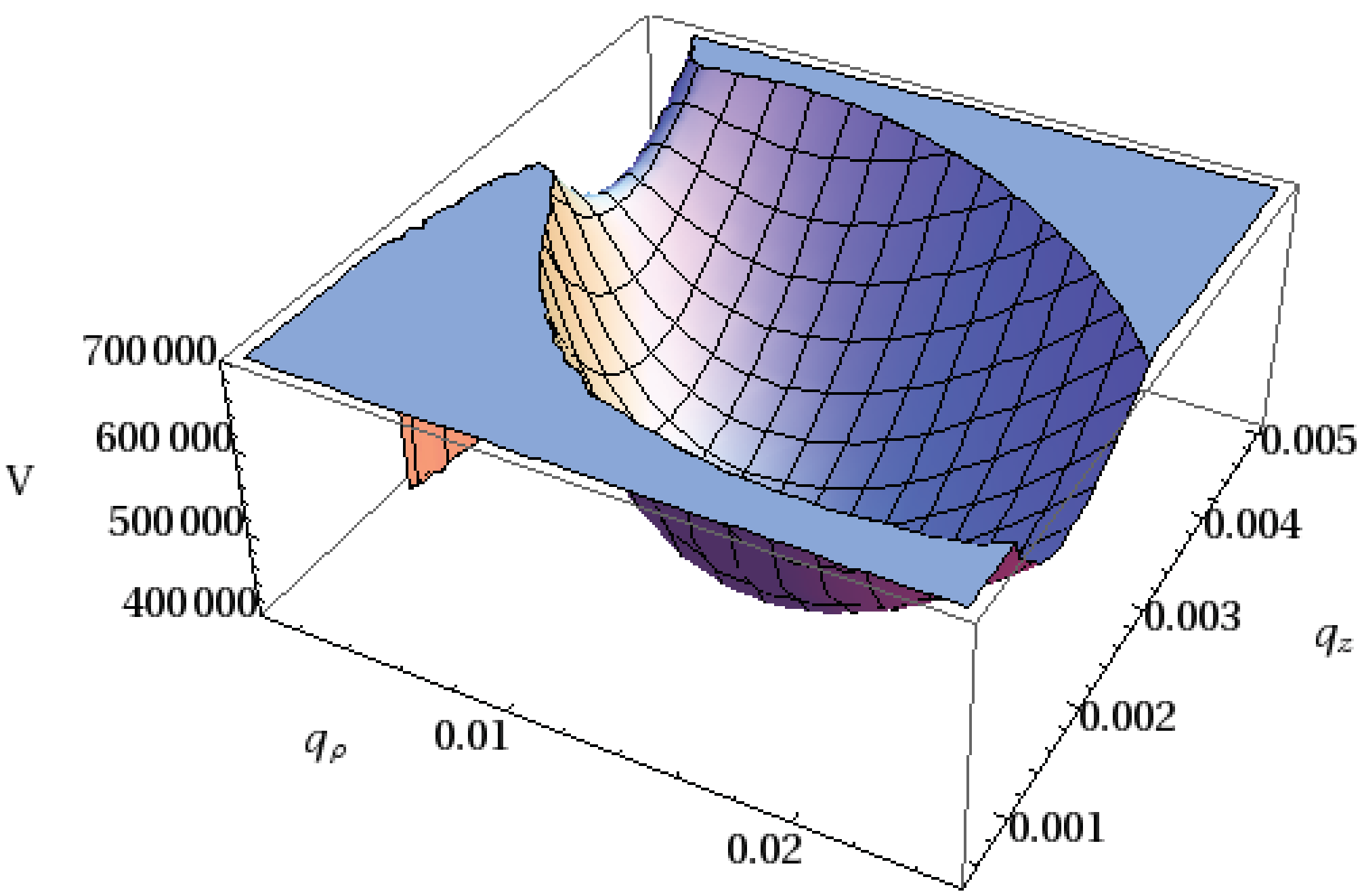}
\end{figure}

We choose a Gaussian trial wave function adapted to the axisymmetric 
trap geometry,  $\psi(\rho,z,t) = e^{i(A_\rho \rho^2+ A_zz^2+\gamma)}$, where
the  complex width parameters $A_\rho$, $A_z$, and the complex phase
are functions of time. Their dynamical equations follow from the time-dependent
variational principle  (\ref{vareq}). Introducing new variables 
$q_\rho, q_z, p_\rho, p_z$ via
\begin{equation}
{\rm Re}\; A_\rho = {p_\rho}/{(4 q_\rho)},~~
{\rm Im}\; A_\rho = {1}/{(4 q_\rho^2)},~~
{\rm Re}\; A_z = {p_z}/{(4 q_z)},~~
{\rm Im}\; A_z = {1}/{(8 q_z^2)}
\end{equation}
one finds that their dynamical equations
are equivalent to the canonical equations of motion belonging to the
Hamiltonian 
\begin{eqnarray}\label{hamiltoniandd}
H &=& T+V =\frac{p_\rho^2}{2}+\frac{p_z^2}{2} \; +\frac{1}{2 q_\rho^2}+  2 \gamma_\rho^2 q_\rho^2+\frac{a/a_{\rm d}} {2 \sqrt{2\pi} q_\rho^2q_z} + \frac{1}{8 q_z^2}+2 \gamma_z^2 q_z^2\nonumber \\
&+&\frac{1+{q_\rho^2}/{q_z^2}-{3 q_\rho^2 \arctan \sqrt{{q_\rho^2}/{(2 q_z^2)}-1}}{\Big/}{\left(q_z^2 \sqrt{{2 q_\rho^2}/{q_z^2}-4}\right)}}{6 \sqrt{2 \pi} q_\rho^4 q_z\left({1}/{q_z^2}-{2}/{q_\rho^2}\right)}\,.
\end{eqnarray}
Thus the variational ansatz has turned the problem into one corresponding to 
a two-dimensional nonintegrable Hamiltonian system, which will exhibit all 
the features familiar from nonlinear dynamics studies of such systems. From 
the shape of the potential, which is shown in 
Fig.~\ref{vnumplot2} as a function of the "position'' variables
$q_\rho, q_z$, these features can  already  be read  off  qualitatively.
At the potential minimum sits the stable 
stationary ground state (elliptic fixed point), while at the saddle point 
one finds an unstable excited stationary state (hyperbolic fixed point).
To quantitatively characterize the dynamics of the variational condensate wave functions 
 we follow the trajectories  in the four-dimensional configuration space spanned 
by the coordinates of the real and imaginary parts of $A_\rho$ and $A_z$.
Since the total mean-field energy is a constant of motion the
trajectories are confined to three-dimensional hyperplanes, and their
behavior can  most conveniently be visualized by two-dimensional 
Poincar{\' e} surfaces of section defined by requiring one of the 
coordinates to assume a fixed 
value. 

\begin{figure}
\centering\includegraphics[width=1\textwidth]{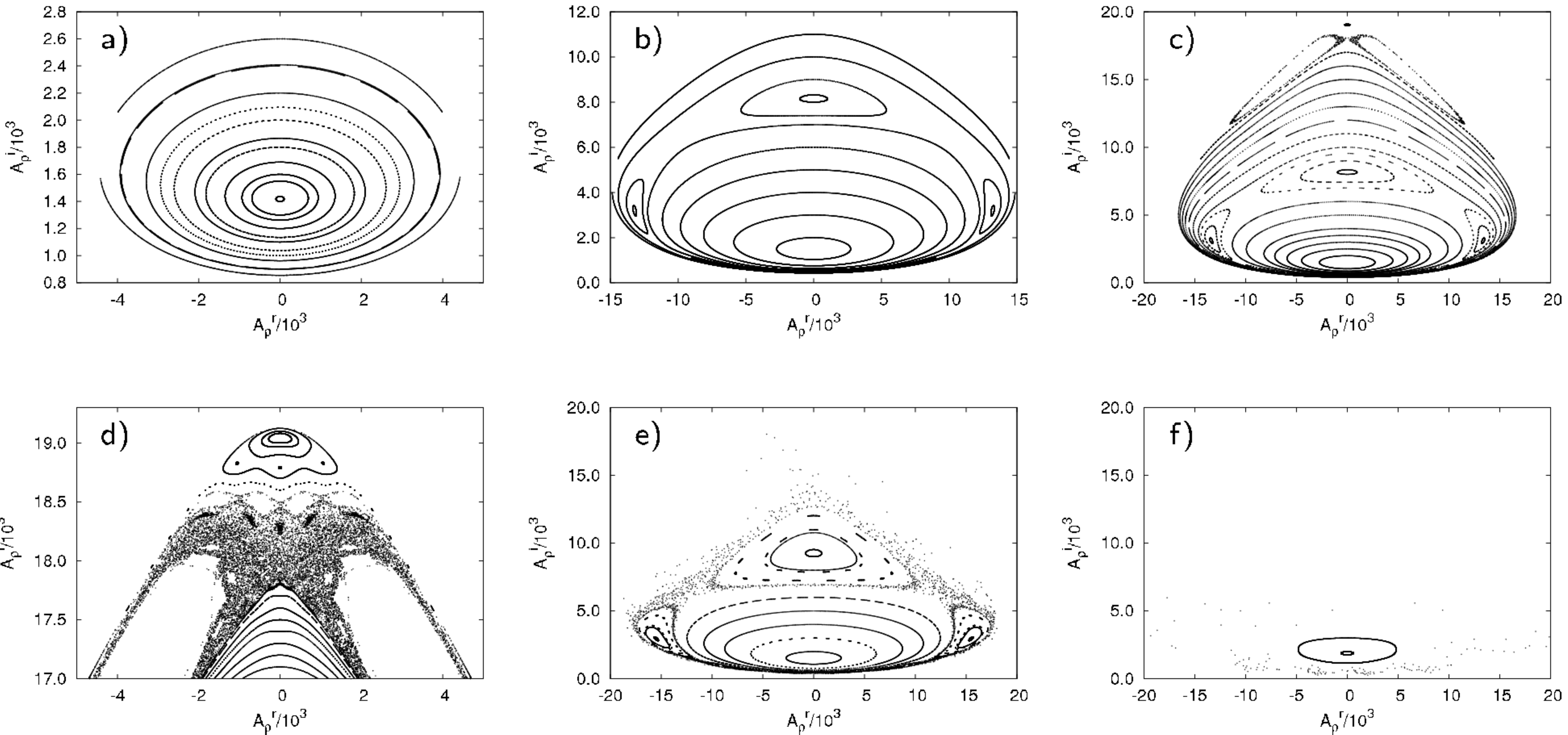}
\caption{Poincar{\' e} surfaces of section of the condensate wave functions 
for dipolar interaction represented by their width 
parameters at the scaled trap frequency $N^2 \bar{\gamma} = 3.4\times 10^{4}$, 
 aspect ratio $\lambda =6$, 
and the scattering length $a/a_{\rm d}=0.1$. The surfaces of section correspond to increasing values
of the mean-field energy (in units of $E_{\rm d}$): a) $NE = 4.5 \times 10^5$, b) $NE =  6.00 \times 10^5$, c) and d)  $NE =  6.24 \times 10^5$,
e)  $NE =  9 \times 10^5$, f)  $NE =  6.00 \times 10^6$.
} 
\label{poincare}
\end{figure}

We  consider
Poincar{\' e} surfaces of section defined by the condition that 
the imaginary part of  $A_z(t)$ is zero. Each time the trajectory 
 crosses the plane ${\rm Im}\;{A_z} = 0$, the 
 real and  imaginary
parts of $A_\rho(t) = A_\rho^r(t) + i A_\rho^i(t)$ are recorded. 
In Fig.~\ref{poincare} surfaces of section  are plotted for 
five different, increasing, values of
the mean-field energy. 
The physical parameters of the experiment of Koch
et al. \cite{koch08}  are adopted, and the scattering length is fixed to 
$a/a_{\rm d}=0.1$, away from its critical
value.
At these parameters, the variational mean-field energy of the ground state 
is $N E_{\rm gs} =4.24 \times 10^5$ (in units of $E_{\rm d}$) and represents the local minimum on the two-dimensional
mean-field energy landscape, plotted as a function of the width parameters. The variational energy
of the second, unstable, stationary state at these experimental parameters is $N E_{\rm es}=6.24 \times 10^5$, it corresponds to 
the saddle point on the mean-field energy surface. 
Between these two energy values the motion on the trajectories is 
bound, while for energies above the saddle-point energy
the motion on the trajectories can become unbound: once the saddle point is traversed by 
a trajectory $A_\rho(t)$, $A_z(t)$, the parameters run to infinity, 
meaning a shrinking of the quantum state to vanishing width, i~e., a collapse of the condensate takes place.

The energy in Fig.~\ref{poincare} a) lies slightly above the energy of the stationary ground state.
The initially stationary state has evolved into a periodic orbit (fixed point in the surface of section), 
corresponding to
a state of the condensate whose motion  is periodic. The oscillations of the width parameters
$A_\rho(t)$ and $A_z(t)$ represent oscillatory stretchings of the condensate along the $\rho$ and $z$
directions.  The  stable {\em periodic} 
orbit in the surface of section is surrounded by elliptical, quasi-periodic orbits, representing quasi-periodic
oscillations of the condensate. As the energy is increased further, in Fig.~\ref{poincare} b), new periodic
and quasi-periodic orbits are born, and the motion is still regular.  In Fig.~\ref{poincare} c) we have reached 
the saddle-point energy. Now chaotic orbits have appeared in the vicinity of the unstable excited stationary state (hyperbolic fixed point). Figure~\ref{poincare} d) shows an enlargement of this 
region in phase space.
In contrast to the (quasi-) periodic stretching oscillations 
of the condensate within the elliptical islands, the chaotic motion of the parameters describes
a condensate which does not yet collapse but whose widths  fluctuate irregularly.

In the surfaces of section shown in Fig.~\ref{poincare} e) and f), with mean-field energies well above the saddle-point energy,   
regular islands are still clearly visible.
These stable islands are surrounded by chaotic trajectories. 
Since ergodic motion along  these trajectories comes 
close to every point in the configuration space, the chaotic motion sooner or later leads to a crossing of 
the saddle point and then to the collapse of the condensate wave functions.
It can be seen that with growing energy  
above the saddle point  the sizes of the stable regions shrink. The kinematically 
allowed regions surrounding
the stable islands are hardly recognizable any more since  high above 
the saddle-point energy the 
chaotic motion becomes more and more unbound, and thus trajectories cross the Poincar{\' e} surfaces
of section only a few times, if ever, before they escape to infinity and  collapse takes place.

It must be stressed, however, that stable islands persist even far above the saddle-point energy, 
implying the existence of quasi-periodically oscillating  nondecaying modes of dipolar condensate wave functions.

\section{Summary and Conclusions}
We have demonstrated that variational forms of the Bose-Einstein condensate wave functions
convert the condensates via the Gross-Pitaevskii equation into Hamiltonian systems that can 
be studied using the methods of nonlinear dynamics. We have also shown that these results
serve as a useful guide in the search for nonlinear dynamics effects in the numerically
accurate quantum calculations of Bose-Einstein condensates with long-range interactions.
The existence of stable islands as well as chaotic regions for excited states of 
dipolar Bose-Einstein condensates is a result that could be checked experimentally. 
One way of creating the collectively excited states one might think of  is
to prepare the condensate in the ground state, and then to non-adiabatically reduce the trap frequencies.

One might question
whether the Gross-Pitaevskii equation is adequate to describe the types
of complex dynamics 
discussed in this paper in ``real'' condensates. For example, in the chaotic regime 
local density maxima might occur for which losses by two-body  or three-body collisions 
would have to be taken into account. However, by virtue of the scaling laws 
discussed in Sections  \ref{scaling1/r}  and  \ref{scaling}
parameter ranges can always be found where the particle densities remain small even in 
these regimes, and the Gross-Pitaevskii equation is applicable.

The advantage of the simple variational ansatz  adopted in this paper is that
the analysis of the nonlinear dynamical properties of Bose-Einstein condensates 
becomes particularly transparent. Numerical quantum calculations to confirm the variational 
findings for dipolar gases and the extensions to structured condensate states 
\cite{goral00,dutta07} are under way. 
We have already seen in Sec.~\ref{1/r}, by 
comparing variational and accurate numerical quantum results for 
 Bose-Einstein  condensates with attractive
$1/r$ interaction, that  the nonlinear dynamical properties predicted 
by the variational calculation were confirmed by the full quantum calculations. 
We therefore have good reason to believe that this will also be true 
once the full quantum calculations of the dynamics of excited condensate 
wave functions of  dipolar gases have become available.


\begin{theacknowledgments}
Part of this work has been supported by Deutsche Forschungsgemeinschaft. H.C. 
is grateful for support from the Landesgraduiertenf\"orderung of
the Land Baden-W\"urttemberg.
\end{theacknowledgments}



\bibliographystyle{aipproc}   


\end{document}